\documentclass[twocolumn,preprintnumbers,amsmath,amssymb,floatfix,superscriptaddress,nofootinbib]{revtex4}

\usepackage{amsmath,hyperref,amssymb}
\usepackage{graphicx}
\usepackage{dcolumn}
\usepackage{bm}
\usepackage{verbatim}
\usepackage{tabularx}
\usepackage{slashed}
\usepackage{cancel}
\usepackage[boxsize=2em,aligntableaux=center]{ytableau}
\usepackage{float}

\def\be{\begin{equation}}
\def\ee{\end{equation}}
\def\bea{\begin{eqnarray}}
\def\eea{\end{eqnarray}}

\begin{document}

\vspace*{-30mm}

\title{Anomalous solutions to the strong CP problem}

\author{Anson Hook}
\affiliation{School of Natural Sciences, Institute for Advanced Study, Princeton, New Jersey 08540, USA}

\vspace*{1cm}

\begin{abstract} 

We present a new mechanism for solving the strong CP problem using a $\mathbb{Z}_2$ discrete symmetry and an anomalous $U(1)$ symmetry.  A $\mathbb{Z}_2$ symmetry is used so that two gauge groups have the same theta angle.  An anomalous $U(1)$ symmetry makes the difference between the two theta angles physical and the sum unphysical.  Two models are presented where the anomalous symmetry manifests itself in the IR in different ways.  In the first model there are massless bifundamental quarks, a solution reminiscent of the massless up quark solution.  In the IR of this model, the $\eta'$ boson relaxes the QCD theta angle to the difference between the two theta angles - in this case zero.  In the second model, the anomalous $U(1)$ symmetry is realized in the IR as a dynamically generated mass term that has exactly the phase needed to cancel the theta angle.  Both of these models make the extremely concrete prediction that there exist new colored particles at the TeV scale.

\end{abstract}

\maketitle

The smallness of QCD's $\overline \theta$ angle has been a mystery for many years.  The physically observable angle is 
\bea
\overline \theta = \theta + \text{arg} \, \text{det} Y_u Y_d
\eea
where $\theta$ is the theta angle and $Y_{u,d}$ are the up and down type yukawas respectively.  Measurements show that $\overline \theta$ must be smaller than $10^{-10}$~\cite{Baker:2006ts}.  This result is especially surprising considering that the CP violating phase in the CKM matrix is order one.  As both CP violating phases have contributions from the yukawa matrices, it is surprising that one should be large while the other is so small.  This difference is a fine tuning of ten orders of magnitude and begs a dynamical explanation.

There are two broad categories of solutions to the strong CP problem.  The first type are solutions based on the CP and P discrete symmetries.  The solutions which use the CP symmetry start with a CP invariant theory and spontaneously break it in such a way that the theta angle vanishes at tree level while the CKM phase is large.  The most well known of these types of theories is the Nelson-Barr mechanism~\cite{Nelson:1983zb,Barr:1984qx}.  Other solutions based off of P involve doubling the matter content of the SM such that the opposite parity sector carries the opposite theta angle\cite{Barr:1991qx}.  Diagonal subgroups thus have non-zero CKM phases but vanishing $\overline \theta$. 

The second class of solutions are based off of anomalous symmetries.  The idea behind these solutions is that in the UV there exists an anomalous symmetry which can be used to rotate away the theta angle and render it unphysical.  These solutions are differentiated from each other by how the anomalous symmetry is realized in the IR.  One popular IR realization of the anomalous symmetry is the  axion~\cite{Peccei:1977hh,Peccei:1977ur,Weinberg:1977ma,Wilczek:1977pj}.  In the axion solution to the strong CP problem, an anomalous symmetry is spontaneously broken yielding a pseudo goldstone boson, the axion.  QCD dynamics generate a potential for the axion.  At the minimum of the potential, the axion vev cancels $\overline \theta$.

Another solution to the strong CP problem that uses an anomalous symmetry is the massless up quark solution~\cite{'tHooft:1976up}, which is currently disfavored by data~\cite{Beringer:1900zz}.  In the presence of a massless up quark, a chiral rotation of the up quark can remove $\theta$ without changing any physical parameters of the theory.  $\theta$ is thus an unobservable parameter.  After confinement, there should be another dual description of how $\theta$ is removed.  This dual description is accomplished by the $\eta'$ boson.  In the large N limit, the $\eta'$ boson has a small mass and can be incorporated into the low energy effective field theory of the goldstone bosons.  The low energy effective description for the $\eta'$ boson\cite{DiVecchia:1980ve} is
\bea
\mathcal{L} = f_\pi^2 \text{Tr} \partial_\mu \Sigma \partial^\mu \Sigma^\dagger + a f_\pi^3 \text{Tr} \left (m \Sigma + m^\dagger \Sigma^\dagger \right ) \\
\nonumber
+ b f_\pi^4 \left (\theta + \frac{i}{2} (\text{Tr} \log \Sigma  - \text{Tr} \log \Sigma^\dagger) \right )^2 + \cdots
\eea
where $a$ is $\mathcal{O}(1)$, $b$ is $\mathcal{O}(1/N)$ and $m$ is the mass of the quarks.  $\Sigma$ is the non-linear sigma field describing the breaking of $U(3)_L \times U(3)_R$ down to the diagonal, i.e. it contains the $\eta'$ boson in addition to the usual pions.  The $\eta'$ vev is stabilized around the theta angle.  All additional higher dimensional operators are a function of $\eta'-\theta f_\pi$ as required by the anomalous symmetry.  Thus once $\eta'$ is integrated out, the only place in the Lagrangian where $\theta$ appears is in the mass terms.  
If the mass is zero then the entire IR Lagrangian is independent of $\theta$.  This is the IR description of the massless up quark solution to the strong CP problem.

In the absence of an up quark mass, the $\eta'$ boson has a shift symmetry that relaxes the $\theta$ angle to 0.  The difference between the massless up quark solution and the axion solution is that the observed $\eta'$ boson obeys
\bea
m_{\eta'}, f_{\eta'} \approx \Lambda_{QCD}
\eea
while the unobserved axion typically has
\bea
m_a f_a = f_\pi m_\pi \frac{\sqrt{m_u m_d}}{m_u + m_d}
\eea
Both of these solutions are realized in the IR as scalars with a shift symmetry that renders theta unphysical.

In this work, we consider a using a $\mathbb{Z}_2$ discrete symmetry in conjunction with an anomalous symmetry to solve the strong CP problem.  The $\mathbb{Z}_2$ discrete symmetry takes the Standard Model (SM) to another mirror copy of the SM\footnote{For a mirror world axion based solution to the strong CP problem see~\cite{Berezhiani:2000gh}.}.  Massless quarks are introduced in a manner such that there is an anomalous $U(1)$ symmetry that results in the sum of the two theta angles being unphysical while the difference is physical.  The presence of a mirror copy of the SM allows for several unique methods to dynamically remove the massless quarks from the IR.

We now present a simple theory which uses a $\mathbb{Z}_2$ discrete symmetry and an anomalous symmetry to obtain a vanishing $\overline \theta$.  We first start off with two copies of the SM related by a $\mathbb{Z}_2$ symmetry.  For notational convenience, the mirror SM will have all of its fields and gauge groups primed.  Because of the $\mathbb{Z}_2$ symmetry, the two theories have equal theta angles and yukawa matrices.  The $\mathbb{Z}_2$ symmetry is spontaneously broken in such a way that the vev of the mirror Higgs field is much larger than the vev of our Higgs field.  The mirror Higgs obtains a large ``natural" vev while our Higgs obtains a small vev.  The large ratio of the vevs is the hierarchy problem, which we do not address in this paper.  How the $\mathbb{Z}_2$ symmetry is spontaneously broken is unimportant for the solution to the strong CP problem but will become important in the context of higher dimensional operators as will be discussed later.

After the two Higgses, $H$ and $H'$, obtain their different vevs, the $\mathbb{Z}_2$ symmetry is broken and the two theta angles are no longer required to be equal.
RG flow from $\langle H' \rangle$ to $\langle H \rangle$ generates a non-zero difference in the theta angles that is much smaller than the $10^{-10}$ experimental bounds\cite{Ellis:1978hq}.  For all intents and purposes we can treat the two sectors as having identical theta angles.  Any subtlety regarding the difference of the two theta angles can be removed completely by considering a supersymmetric model where the theta angle does not run. 

There are two simple ways in which an anomalous symmetry can be added to the picture.  The simplest way is to add a pair of massless bifundmentals $\psi_B$ and $\overline \psi_B$ that are charged under $SU(3)_c$ and $SU(3)_{c'}$.  It is simple to see that using the anomalous rotation of $\psi_B$ and $\overline \psi_B$ that the two theta angles can be simultaneously set to zero.  After $H'$ obtains a large vev, all of the dual matter aside from $\psi_B$ and $\overline \psi_B$ becomes massive.  The $SU(3)_{c'}$ gauge group has only three flavors and confines.  Due to chiral symmetry breaking, there is an octet of scalars.  The octet of scalars is an octet under $SU(3)_c$ and obtains a mass from gauge boson loops in much the same way that $\pi^\pm$ obtains a larger mass than $\pi^0$ from loops involving the photon.  These loops give the scalars a mass a factor of few below the mass of the $\rho'$ mesons of the $SU(3)_{c'}$.

Like the case with massless quarks, what is important for the low energy dynamics is how the $\eta'$ boson behaves and not the dynamics of any of the pseudo goldstone bosons.  
To the extent that $N=3$ can be approximated by the large N limit, we can write an effective low energy theory of the $\eta'$\cite{DiVecchia:1980ve} which is
\bea
\label{Eq: chiral}
\mathcal{L} = \frac{g^2}{32 \pi^2} \left ( \theta - \frac{\eta'}{f_{\eta'}} \right ) F \tilde F + \frac{m_{\eta'}^2}{2} \left ( \eta' - f_{\eta'} \overline \theta' \right )^2 + \cdots
\eea
where the theta angles appear in a manner required by the anomalous $U(1)$ symmetry.  For simplicity's sake we do not include the non-linear sigma field $\Sigma$.
Unlike the axion, the mass of the $\eta'$ is not set by the mass of the quarks but instead by the topological charge density\cite{Witten:1979vv}.
\bea
m_{\eta'}^2 = \frac{4 N_f}{f_\pi^2} \frac{d^2 E}{d\theta^2}|_{\overline \theta'=0, \text{no quarks}} \approx \Lambda_{QCD'}^2
\eea
where $N_f=3$ is the number of flavors, $f_{\eta'} \approx f_\pi$ is the pion decay constant, and E is the vacuum energy density evaluated when $\overline \theta'=0$ and in the absence of quarks.  From Eq.~\ref{Eq: chiral}, we see that after integrating out $\eta'$ that the $SU(3)_c$ theta angle becomes $\theta - \overline \theta' = - \text{arg} \, \text{det} Y_u Y_d$.

In the IR, the $SU(3)_c$ has a boundary condition at $\Lambda_{QCD'}$ where its $\overline \theta$ is equal to zero.  This solution is like the massless quark solution to the strong CP problem where the $\eta'$ dynamics are important.  In contrast, in composite axion solutions to the strong CP problem it is the dynamics of a pseudo-goldstone boson that obtains a mass from the QCD scale and not the QCD$'$ scale which is important.

There is a second way in which an anomalous symmetry can be added to the theory with two copies of the SM.  We consider adding a complex scalar $\phi$, $N_f$ vector-like flavors of fundamentals ($\psi_Q^i$, $\overline \psi_{Q,i}$) for $SU(3)_c$ and $N_f$ vector-like flavors of fundamentals ($\psi'^i_Q$ and $\overline \psi'_{Q,i}$) for $SU(3)_{c'}$.  The flavors of quarks are exchanged under the $\mathbb{Z}_2$ symmetry while $\phi$ is odd under the symmetry.  We require that the Lagrangian have an anomalous symmetry under which the flavors $\psi$ have charge 1 and the scalar $\phi$ has charge -2.  As before, this anomalous symmetry renders the sum of the two theta angles unphysical.

After $H'$ obtains a vev, all of the dual matter aside from $\psi'^i_{Q}$, $\overline \psi'_{Q,i}$, the $SU(3)_{c'}$ and the $U(1)_{EM'}$ gauge groups are massive.  The $SU(3)_{c'}$ gauge group confines with
\bea
\label{Eq: condensate}
\langle \psi'^i_Q \overline \psi'_{Q,j} \rangle \sim \Lambda_{QCD'}^3 e^{i \overline \theta'/N_f} \delta^i_j
\eea
Because all of the matter content of the mirror has been integrated out, the theta angle of the $SU(3)_{c'}$ gauge group is $\overline \theta' = \theta + \text{arg} \, \text{det} Y_u Y_d$.  Using a chiral rotation, we see that $\overline \theta'$ appears in the quark condensate as shown in Eq.~\ref{Eq: condensate}.  Alternatively, one can note that the $\eta'$ boson is stabilized at $\langle \eta' \rangle = \overline \theta' f_\pi$.  As mentioned before, even after the $\mathbb{Z}_2$ symmetry breaking, $\overline \theta' = \overline \theta$ to very high accuracy.  

The most general potential consistent with the anomalous symmetry for the scalar $\phi$ and the massless quarks is
\bea
V_\phi &=& y e^{i \theta_\phi} \phi (\psi^i_Q \overline \psi_{Q,i} - \psi'^i_Q \overline \psi'_{Q,i}) \\ \nonumber
&+& m^2 \phi \phi^\dagger  + \lambda (\phi \phi^\dagger )^2 + c.c.
\eea
where we have taken y real.  After the $\psi'_Q$ quarks condense, we see that $\phi$ obtains a vev
\bea
\langle \phi \rangle = y \frac{\Lambda_{QCD'}^3}{m^2} e^{- i (\overline \theta'/N_f + \theta_\phi) } + \cdots
\eea
We have made the simplifying assumption that $\lambda \ll 1$ and $m \gg \Lambda_{QCD'}$.  Using a phase rotation on $\phi$ we can see that the phase is exactly correct.
This vev gives a mass $\sim e^{- i \overline \theta' / N_f}$ to the fermions $\psi^i_Q$ and $\overline \psi_{Q,i}$.  After integrating out these quarks, we find that the $SU(3)_c$ gauge group has $\theta = - \text{arg} \, \text{det} Y_u Y_d$.  Thus the invariant theta angle of $SU(3)_c$ is zero.  The IR manifestation of the anomalous symmetry is a dynamically generated mass term that cancels the theta angle!

In the previous calculation, we ignored the back reaction of the vev of $\phi$ on $\langle \psi'^i_Q \overline \psi'_{Q,j} \rangle$ by taking $m$ to be large.  To take into account back reaction, we use the parameterization
\bea
\langle \psi'^i_Q \overline \psi'_{Q,j} \rangle = \Lambda_{QCD'}^3 e^{i \overline \theta'/N_f} \delta^i_j f(y \phi)
\eea
where $f(y \phi)$ is a real function that parameterizes the effects of a non-zero quark mass on the fermion bilinear.  If the quarks were massless, then we know that they confine so that $f(0) \approx 1$.  On the other hand, if the mass of the quarks is larger than $\Lambda_{QCD'}$, then the quarks do not confine and we have $f(m \gtrsim \Lambda_{QCD'}) \approx 0$.  To find the vev of $\phi$, we need to solve the equations of motion which are approximately
\bea
m^2 \phi - y e^{-i \theta_\phi} \Lambda_{QCD'}^3 e^{-i \overline \theta'/N_f} f(y \phi) = 0
\eea
From the various limits of $f(y \phi)$, we see that $y \langle \phi \rangle \lesssim \Lambda_{QCD'}$ so that the mass of $\psi^i_Q$ and $\overline \psi_{Q,i}$ is $\lesssim \Lambda_{QCD'}$. 

We now consider the effect of higher dimensional operators on these solutions to the strong CP problem.  There are two types of operators which can cause an effect.  The first kind are higher dimensional operator which break the anomalous symmetry used to make the sum of the theta angles unphysical.  As we have learned from string theory UV completions of axion models, there are good reasons to expect that quantum gravity effects which break the anomalous symmetries are greatly suppressed.
The second more dangerous type of higher dimensional operators respect all symmetries.  Dangerous operators include
\bea
\frac{g^2}{32 \pi^2} \left ( \frac{H H^\dagger}{M_{pl}^2} G \tilde G + \frac{H' H'^\dagger}{M_{pl}^2} G' \tilde G' \right )
\eea 
If the pre-factor was not included, then if one used a chiral rotation to rotate this coupling into the yukawa couplings, one would find that the operator is not suppressed by $M_{pl}^2$ times an order one constant but instead by an order $32 \pi^2/g^2$ constant.  Thus the pre-factor is needed if higher dimensional operators are to be suppressed by the planck scale and not by a parametrically lower scale.
After the $\mathbb{Z}_2$ symmetry is broken, this operator causes the difference between the theta angles to be non-vanishing.  In order not to reintroduce the strong CP problem, this operator requires that the vev of $H'$ be smaller than $\sim 10^{14}$ GeV.

Depending on how the $\mathbb{Z}_2$ is broken, there may be dimension 5 operators which impose stronger constraints.  If the $\mathbb{Z}_2$ symmetry is broken by a real scalar $\Phi$ odd under the discrete symmetry\cite{Chang:1984uy}, then there is a dangerous dimension 5 operator
\bea
\frac{g^2}{32 \pi^2} \frac{\Phi}{M_{pl}} \left ( G \tilde G -  G' \tilde G' \right )
\eea
If the only fine-tuning allowed in the theory is that needed to solve the hierarchy problem, then this operator leads to the constraint that $\Phi \sim H'$ be smaller than $10^9$ GeV, which results in new colored states with mass below 100 GeV and is experimentally ruled out.  Thus we require that the $\mathbb{Z}_2$ is broken in such a way that only dimension 6 operators can be written.  The simplest way to avoid this problem is to instead introduce a scalar $\Phi$ which is a doublet under a new $SO(2)$ gauge symmetry and impose that it transform as $\Phi \rightarrow i \Phi$ under the discrete $\mathbb{Z}_2$ symmetry.  In this way, the quartic $\Phi^2 ( H H^\dagger -  H' H'^\dagger)$ can be tuned against the mass term to result in a very light Higgs and no dangerous dimension 5 operators can be written.

\begin{figure}[!t]\begin{center}
\includegraphics[width=0.45\textwidth]{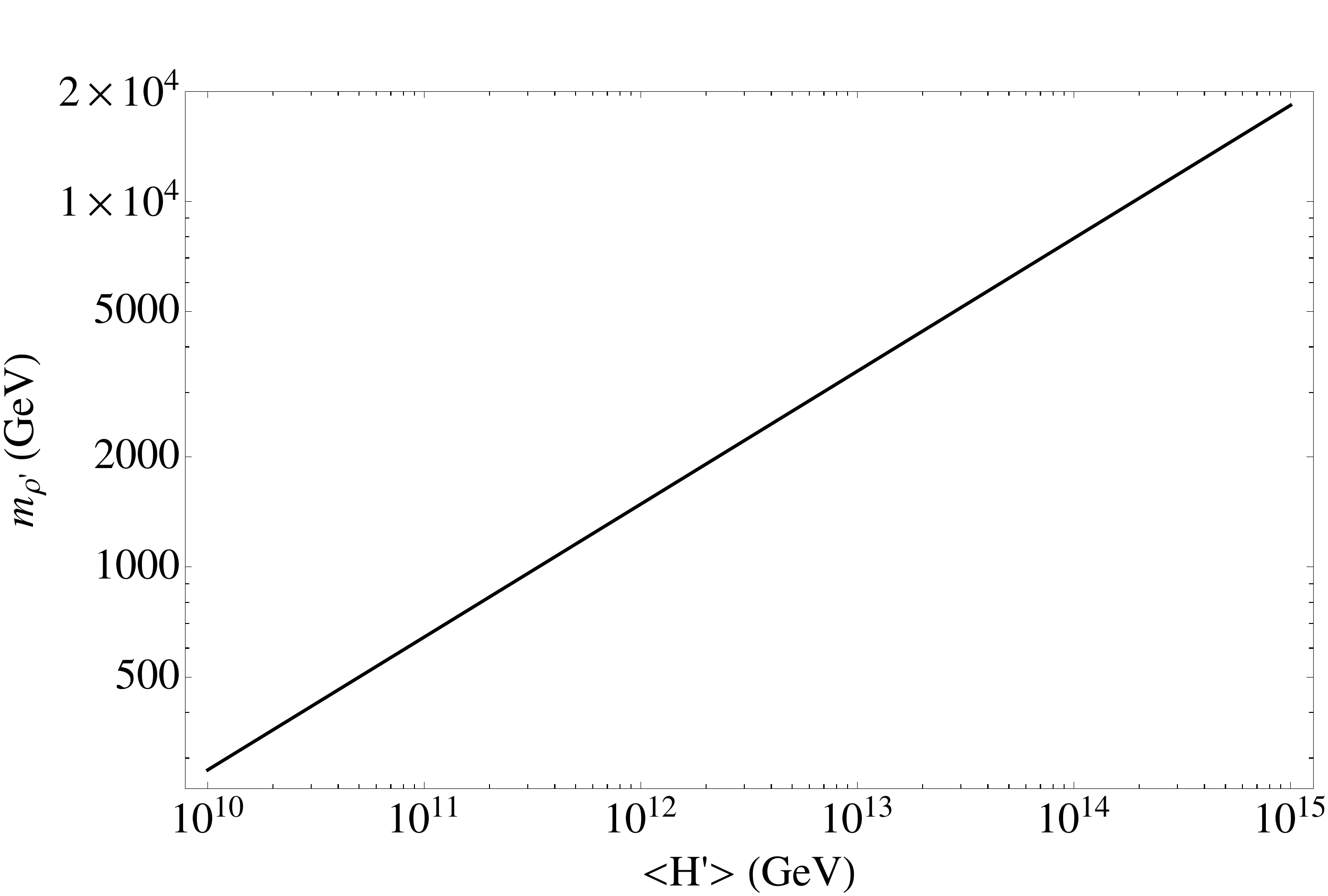}\end{center}
\caption{The mass of the $\rho'$ meson in the mirror sector as a function of the mirror Higgs vev.  The mass of the scalar color octet will be a factor of few below the mass of the $\rho'$.}
\label{Fig: instability}
\end{figure}
The phenomenological consequences of these solutions are tied to the strong coupling scale of the mirror sector.  In the solution involving massless bifundamentals, the first signatures would be the color octet bosons.  The color octet obtains a mass
\bea
m_{\pi'}^2 \approx \frac{3 \alpha_s C_2(\text{adj})}{4 \pi} m_{\rho'}^2
\eea
at 1-loop which is about a factor of 4 below the mass of the $\rho'$ gauge bosons.  
In Fig.~\ref{Fig: instability} we plot the mass of the $\rho'$ meson (assuming that $m_{\rho}/\Lambda_{QCD} = m_{\rho'}/\Lambda_{QCD'}$) as a function of the vev of the mirror Higgs.

\begin{figure}[!t]\begin{center}
\includegraphics[width=0.45\textwidth]{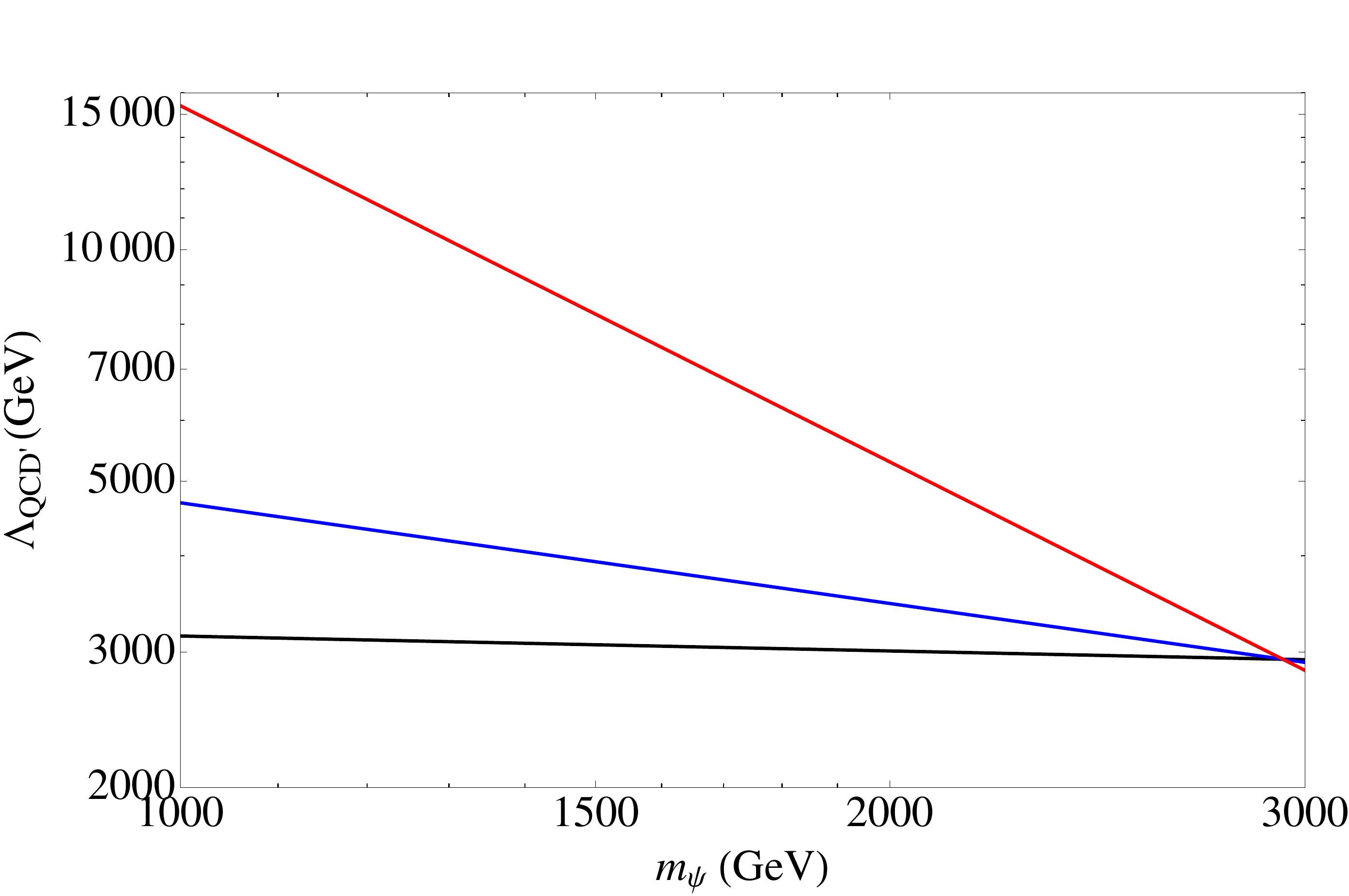}\end{center}
\caption{$\Lambda_{QCD'}$ as a function of the mass of the $N_f$ flavors of quarks $\psi^i_Q$.  The vev of the mirror Higgs is set to be $10^{14}$ GeV.  The mass of the $N_f$ flavor of quarks is bounded by about 3 TeV.  The black, blue and red lines are $N_f = 1,5$ and 10 respectively.}
\label{Fig: instability2}
\end{figure}

In the second solution involving a scalar and fundamentals, the first signature would be $N_f$ vector-like fundamental quarks with mass $\lesssim \Lambda_{QCD'}$.
 In Fig.~\ref{Fig: instability2}, we plot $\Lambda_{QCD'}$ as a function of the mass of the $N_f$ flavors of fundamental quarks setting the vev of the mirror Higgs to $10^{14}$ GeV.  We see that there is an upper bound on the mass of these new quarks which is about 3 TeV.

The constraint from higher dimensional operators on the vev of $H'$ and the experimental bounds coming from the search for colored particles leaves open a very interesting region of parameter space.  Due to higher dimensional operators, we have
$\langle H' \rangle < 10^{14}$ GeV.
In the theory with a scalar octet, we find that the upper bound on the Higgs vev sets an upper bound of 8 TeV on the mass of the $\rho'$ meson.  The mass of the scalar octet is then only a few TeV and potentially observable at the LHC.  In the theory with additional flavors of quarks, the upper bound on the Higgs vev sets the quarks to be lighter than some order one number times 3 TeV.  

Both of these simple models contain collider observable particles.  Much like how the $\pi^0$ decays into a pair of photons through the Wess-Zumino term in the chiral Lagrangian, the color octet decays into a pair of gluons via the Wess-Zumino term.  
In the other model, the fundamental fermions are stable because they are the lightest particles charged under a vector-like $U(1)$.  

In both of these simple models, there is also a massless $U(1)_{EM'}$ gauge boson.  Due to the structure of both theories, the kinetic mixing between $U(1)_{EM}$ and $U(1)_{EM'}$ is suppressed by many loop factors.  As a result, this new gauge boson is not constrained by current experiments.  Alternatively, the two $\mathbb{Z}_2$ copies of the SM could share a $U(1)_Y$ so that there is only one photon and no additional massless $U(1)_{EM'}$ gauge boson.

Both of these models require that there exist colored particles with mass in the TeV range.  This connection to the LHC is extremely exciting as the LHC is probing this region of parameter space.  The color octets give a 4 jet signal with two resonances.  This search is being done (e.g. see Ref.~\cite{Khachatryan:2014lpa}) and while bounds were not presented for this particular model, a very rough estimate gives a bound of order 500 GeV or so.  On the other hand, the fundamental fermions are collider stable colored particles.  While these fermions have cross sections smaller than gluinos, Ref.~\cite{Chatrchyan:2013oca} demonstrates that the bound on these types of particles is roughly a TeV or so.

In this short note, we have demonstrated a new avenue for making $\overline \theta$ vanish.  This approach uses a $\mathbb{Z}_2$ discrete symmetry and an anomalous symmetry.  We have presented two simple models that illustrates this new approach to the problem.  The key mechanism which allows this approach to work is the observation that confinement of the mirror SM allows for unique ways in which the massless quarks can be removed from the IR.  They can be removed from the IR by confinement or by a dynamically generated mass with exactly the phase needed to cancel $\overline \theta$.

\acknowledgments
We thank Nima Arkani-Hamed for helpful discussions, Gordan Krnjaic and Carlos Tamarit for collaboration in early stages of the project, and Luca Vecchi for comments on the decay of the color octet.  AH is supported by the DOE grant de-sc000998.  AH thanks the Perimeter Institute where this work was initiated.

\bibliography{ref}

\end{document}